\documentclass[letterpaper]{article} % DO NOT CHANGE THIS
\usepackage{aaai25}  % DO NOT CHANGE THIS
\usepackage{times}  % DO NOT CHANGE THIS
\usepackage{helvet}  % DO NOT CHANGE THIS
\usepackage{courier}  % DO NOT CHANGE THIS
\usepackage[hyphens]{url}  % DO NOT CHANGE THIS
\usepackage{graphicx} % DO NOT CHANGE THIS
\usepackage{natbib}  % DO NOT CHANGE THIS AND DO NOT ADD ANY OPTIONS TO IT
\usepackage{caption} % DO NOT CHANGE THIS AND DO NOT ADD ANY OPTIONS TO IT
\usepackage{algorithm}
\usepackage{algorithmic}

\usepackage{booktabs}

\usepackage{newfloat}
\usepackage{listings}
\DeclareCaptionStyle{ruled}{labelfont=normalfont,labelsep=colon,strut=off} % DO NOT CHANGE THIS
\floatstyle{ruled}
\newfloat{listing}{tb}{lst}{}
\floatname{listing}{Listing}
\title{The Folly of AI for Age Verification}
\author{
    Reid McIlroy-Young
}
\affiliations{
    Harvard University\\
    reidmcy@seas.harvard.edu
}

\begin{document}

\maketitle

\begin{abstract}

In the near future a governmental body will be asked to allow companies to use AI for age verification. If they allow it the resulting system will both be easily circumvented and disproportionately misclassify minorities and low socioeconomic status users. This is predictable by showing that other very similar systems (facial recognition and remote proctoring software) have similar issues despite years of efforts to mitigate their biases. These biases are due to technical limitations both of the AI models themselves and the physical hardware they are running on that will be difficult to overcome below the cost of government ID-based age verification. Thus in, the near future, deploying an AI system for age verification is folly.

\end{abstract}

\section{Application Context}

There is a growing demand for systems that can cheaply and efficiently identify a website user's age. \textbf{What does it do?} There are many AI systems available for sale or download under open source licenses that nominally solve the age verification problem by taking a photo of the user and outputting their age in real time. \textbf{What is an application that it might be used for?} These systems are an obvious consideration for companies attempting to meet government mandates without increasing friction for users or requiring handling of Personally Identifiable Information. \textbf{What applications has it been used for?} The use of these systems in practice is highly suspect as they are likely to underperform in real world tests, thus most are used for internal tools and are part of a larger set of tools a company uses for identity checks or age verification. \textbf{What attributes or properties does it have that might make it a good/bad fit for specific public missions?} These systems tend to have systemic biases against under-represented groups, and can be easily circumvented by users via technical or physical means. As such, pure AI-based age verification will not work with current systems, and the fixes needed will require significant effort both by the developers and third parties like webcam manufacturers.

\section{Introduction}

There is a growing governmental interest in requiring websites to verify the age of their users, with the goal of limiting users below a given age from accessing material that has been deemed harmful. This age gating has risen in prominence with the recent mandate in Australia requiring social media sites to block users under 16~\citep{Mcguirk2024}, countrywide. While, at the more local level many (19) US States have instituted bans on displaying obscene content to minors as of late 2024~\cite{freespeechcoalition}. Most of these laws require the use of government issued ID (\textit{e.g.} Driver's License) to verify age, but some also allow for \textit{commercially reasonable methods} which are not fully specified. Thus, it is likely that website developers are already looking at alternative methods to verify age more cheaply and smoothly, while still plausibly providing success. This paper is arguing that `Artificial Intelligence' (AI) based methods using current computer vision technology cannot work for this task, and that there are fundamental technical limitations to computer vision that make age verification with AI that are unsolved despite decades of intense market demand for their solutions. As such it is folly to consider using AI for age verification.

\section{Background}

In 2012 a team from the University of Toronto presented a new approach to image classification, called \textit{AlexNet}. \textit{AlexNet} used convolutional neural networks (CNN) trained with deep learning to vastly outperform the other competitors. This neural network based approach to image classification had become the standard for computer vision AI. Thus, for this paper we will be considering CNN based computer vision systems as they are the most likely to be deployed and are well understood in the literature.

\begin{figure*}[ht]
    \centering
    \includegraphics[width=1\linewidth]{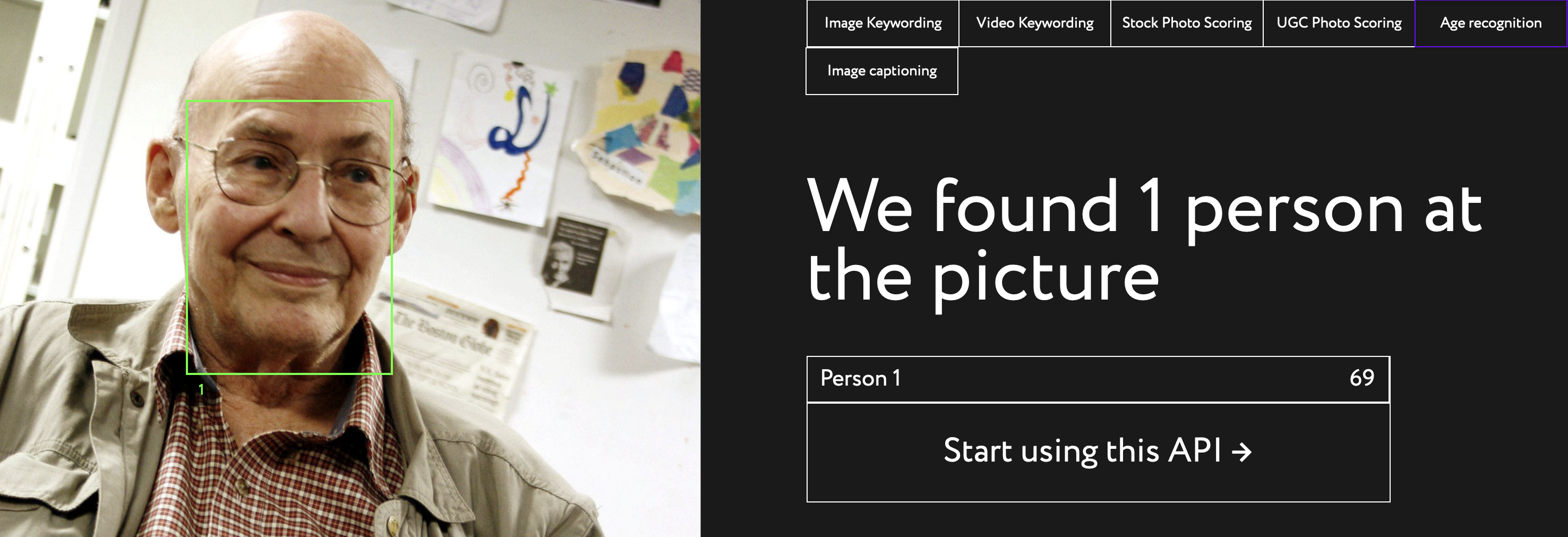}
    \caption{Photograph of Marvin Minsky at age 81, with the output of a commercial age recognition system.}
    \label{fig:marvin}
\end{figure*}

\subsection{Definition of AI}

For this paper we use the term `Artificial Intelligence' (AI) to refer to machine learning models that have been trained via back propagation on some data set. This is a very general definition, and we will mostly treat AI systems as black boxes that take in some input (image) and output a label (age). In practice this means the types of models created using the methods provided in ~\citeauthor{Levi2015AgeAG}~\cite{Levi2015AgeAG}, see figure~\ref{fig:marvin} for an example of one of these systems.

\newpage
\subsection{Fairness Concerns in AI}~\label{sec:fair}

The concerns raised in this paper are based on issues that occur in many other computer vision projects. Here we will focus on computer vision involving recognizing and classifying human faces. The issues of \textit{bias} and \textit{generalization} are much broader and extend far beyond human based tasks. Notably, the slow development of self driving cars is in part due to the problem of `outliers'~\cite{Kirkpatrick2022StillWF} situations that the vision and planning models are not designed for. This is a \textit{lack of generalizability}, the models cannot adapt to new situations. The other main concern we have is of models performing incorrectly on subsets of their inputs, \textit{i.e.} \textit{they have a bias} for or against the subset~\cite{mehrabi2021survey}. An example of this type of bias is a machine learning model that assumes that the word `Programmer' is associated with the word `man'~\cite{bolukbasi2016man}. Importantly the issues of \textit{bias} and \textit{generalization} are tightly coupled, and often co-occur. The solution to both is also well known, but costly, increasing the size and diversity of the training dataset~\cite{buolamwini2018gender} or increasing the computing resources used in running and training the model~\citep{sutton2019bitter}.

\subsubsection{Fairness in Gender Classification}

One of the first studies of bias in image classification was the \textit{Gender Shades}~\cite{buolamwini2017gender,buolamwini2018gender} study. This work looked at commercially available AI systems for gender classification and examined how well they performed on people of different skin tones. The work showed that there was significant drop in performance for darker skinned individuals along with a moderate drop in performance for women. The root cause of this issue is multi-faceted~\cite{Muthukumar2018UnderstandingUG} and work is still ongoing to fully mitigate it~\cite{Laszkiewicz2024BenchmarkingTF}. The current best solution is to use balanced datasets for training~\cite{ChinPurcell2021InvestigatingAD,buolamwini2018gender}, but doing so is expensive and time consuming~\cite{Nadimpalli2022GBDFGB} as such many current systems still fail these tests~\cite{Siddiqui2022AnEO}.

\subsubsection{Remote Proctoring Systems}

A related task to age verification is that of verifying exam integrity where the proctor is only able to observe via remote observation methods, and is limited to matériel the student has at hand. These systems were widely deployed to allow test taking during the COVID-19 pandemic. These remote proctoring systems were both much less effective than in person proctors at preventing cheating~\cite{Newton2023HowCI} and much more stressful for students~\cite{Pokorny2023OutOM}. A major component of these systems is a set of computer vision modules that monitor the test taker, these systems often have issues with darker skinned individuals~\cite{Pokorny2023OutOM, vox2020,Burgess2022WatchingTW}, despite US school system's mandates for equal treatment.

\section{Taxonomy of Concerns for Age Verification AI}

For this analysis we will focus on two categories of potential error. First are \textit{false positives} where the model incorrectly classifies a user as being below the age threshold. Notably, in AI systems these can occur for two different reasons, one is a true false positive where the model makes an incorrect classification given a correctly formatted image, but there is a second case where the model is `uncertain' (\textit{e.g.} given a blurry or dark photo) in this case the developer is also likely to reject the user. The second type of error we will discuss here is \textit{false negatives} where the model incorrectly classifies an underage user as above age.

For this analysis we consider both \textit{false positives} and \textit{false negatives} to be harmful, although the harms from \textit{false positives} will primarily be to the users whereas \textit{false negatives} will primarily be to the website operator and, in theory, society. Table~\ref{tab:taxon} lists the possible sources of error we discuss below along with which case they relate to.

\begin{table*}[t]
    \centering
    \caption{Caption}
    \label{tab:taxon}
    \begin{tabular}{lll}
         Source of Error & Type of Error & Description\\
         \midrule
         USB Webcam Spoofing & \textit{false positive} & User using a USB device to simulate a webcam image\\
         Physical Image Spoofing & \textit{false positive} & User faking parts of the scene visible to the camera\\
         Low Quality Images &\textit{false negative}&Users being unable to provide images of the required standard\\
         Systematic Biases&\textit{false negative}&Models being biased against some set of users\\
         Model Drift & \textit{Both} & Models become less performant and easier to exploit over time\\
    \end{tabular}
    
\end{table*}

\subsection{Image Spoofing}

One possible way to circumvent an computer vision based age verification AI is to manipulate what it sees. Currently there is no cryptographic verification involved in the generic webcam implementation~\cite{msusb}. Thus a precocious individual could \textit{simulate a USB device} either in software or physically that displays synthetic images that will pass the verification. These images can be generated in a variety of ways, notably deepfakes and video game based character simulations can both be done in real time with only moderate consumer hardware requirements~\cite{pashine2021deep}. Notably software and hardware based spoofing are not possible on modern cellphone cameras as they do implement cryptographic verification systems. This type of spoofing attack leads to \textit{false positives}, as it requires active participation by the user.

There are also lower tech ways to spoof images. \textit{Printing images} and placing them in front of the camera is well known to trick for identity verification. This method is unlikely to work on simple age verification system, but more advanced versions like playing a video in-front of the camera or using a 3D mask~\cite{patel2016secure} may work. These are all active methods of spoofing so would lead to \textit{false positives}.

\subsection{Model Errors}

As we discuss in the previous section current facial recognition systems suffer a set of known errors. The first one relevant to age verification is \textit{low performance on low quality images}. The datasets used for training AI systems are often of high quality or with a limited set of possible defects. Thus it takes many iterations to create an AI system that can handle all situations it will encounter when deployed. In practice this means that users whose camera and lighting setups are outside the expected range will have troubles using the system. In practice this will mean that users without the resources to modify the systems will encounter verification issues at a higher rate than those with more resources.

A second major concern is the age verification system being \textit{systematically biased against a specific group}, either due to faults in the training data or biases in how the data are collected~\cite{roth2009looking}. These biases can manifest in the age verification model directly or be part of the facial recognition system that prepares the image for age verification, this is the system that draws the green box seen in figure~\ref{fig:marvin}. Minorities and marginalized groups are likely to be most effected by these biases, and as the biases are based on their identities it would require work by the creators to mitigate the issues.

Finally, there is the generalization problem, models trained today, do not know what tomorrow looks like. Something as simple as a change in hairstyles can lead to changes in model performance~\cite{albiero2021gendered}, so models that deal with people will need to be constantly updated, this change in what AI models are trained on from the real world is called \textit{model drift}. This generalization problem, also cuts both ways, it will slowly reduce model performance in general creating \textit{false negatives}, but also users may also learn strategies to trick it leading to an increase in \textit{false positives}.

\section{Conclusion}

While AI models for age verification exist and nominally perform well if we look at how similar projects have fared in the past we can see that major equity and robustness concerns should be addressed before governments allow them to be used. We see that these AI systems tend to be systematically biased in ways that are endogenous to their training data and due to the physical design decisions made in how AI systems interact with the world. Additionally, we present multiple ways to circumvent these AI models that will need to be addressed so that governments can trust the systems. In conclusion, AI for age verification is folly.

\bibliography{aaai25}

\end{document}